# An Important Decision for Climate Change: Sequestering Carbon in Materials or Underground?

Peter Eisenberger, Columbia University & Equitable Climate Innovations Institute

## Abstract

A first-order analysis concludes it is feasible to store the carbon needed to meet the Paris targets in structural materials and use less energy and at a lower cost than our use of extractive materials, steel, aluminum, and concrete. Switching to a synthetic material industry using $CO_2$ from the air will stimulate economic growth and create increased equity while addressing the threat of climate change. The positive feedback between them will accelerate reaching a global accord to address the dual threats of climate change and equity and may, in fact, be needed to avoid the catastrophic consequences of failing to meet them on time.

## Introduction

Should one sequester the $CO_2$ underground as a waste product as current policy emphasizes or sequester it in valuable construction materials? Our continuing to use extractive materials like iron ore, aluminum ore, and aggregate for construction will involve increasing environmental damage as the demand increases and supplies become more limited. Synthetic carbon fiber and aggregate using $CO_2$ as from the air as the carbon source avoids environmental damage and creates economic value that stimulates economic growth while sequestering the carbon to address the climate change threat. Thus, using synthetic materials to build the needed infrastructure for equity and the climate fight is preferable over deciding who has to pay for treating the $CO_2$ as a waste. Who pays has prevented reaching a global accord that will be easier to reach because fighting the climate threat brings the benefit of equitable economic growth.

 So the above raises three questions:

      1. Will the energy needed to make synthetic materials be less than that using natural resources?

      2. Will the cost of synthetic materials be less than extractive materials?

      3. Will there be enough demand for synthetic materials to sequester the carbon needed for climate protection?

The following analysis is a first-order feasibility analysis of the above questions; it does not predict the future, but it intends to show that the answer to the three questions is yes. Therefore, regarding CO2 as a valuable molecule is better than paying to sequester it as a waste. It does not mean that $CO_2$ should not be sequestered underground as a waste. It should, however, be a backup to converting it into valuable materials and not vice versa as in current policy.

Before answering the three questions, we must recognize that we are undergoing an industrial revolution in transitioning from a natural resource fossil fuel economy to a Renewable Energy and Materials Economy (REME)[1].

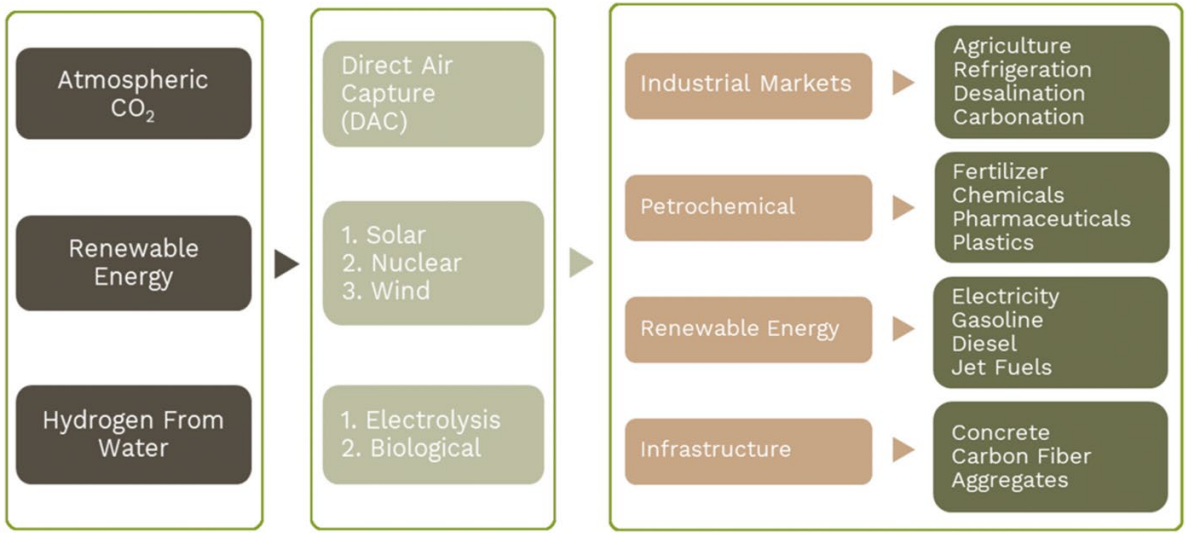

Figure 1. Renewable Energy & Materials Economy - REME mimics Nature

REME creates a closed-cycle carbon economy and provides sequestration. Because it creates economic growth, it will create positive feedback between economic prosperity, climate change protection, and increasing equity[2].

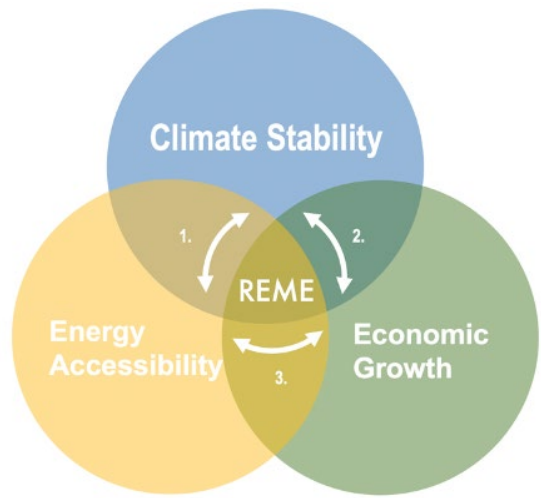

Fiqure 2 - Positive Feedback-1 Increased clean energy accessibility enhances climate stability, 2. Economic growth increases the rate of providing climate stability and the corresponding increase of social peace; 3. The overall increase in energy, including fossil fuels combined with carbon capture and storage during the energy transition, enables enhanced economic growth.

The positive feedback enabled by the REME industrial revolution, changing the negative feedbacks in our current natural resource economy, enables faster global growth rates than predicted by extrapolating our past growth rates into the future. China experienced growth rates of 9% during its industrial revolution, and the US economy grew by 17% annually for the five years we were at war, from 1940 to 1945[3].

We will use an average of 6% for global economic growth for our analysis below and argue that it is conservative once we mobilize to address the dual threats of climate change and inequity. Of course, high growth rates are necessary to achieve global equity between the global north and global south countries, not to mention within each nation. Providing basic needs is measured by the UN Human Development Index is correlated, as shown below with the energy use per capita.

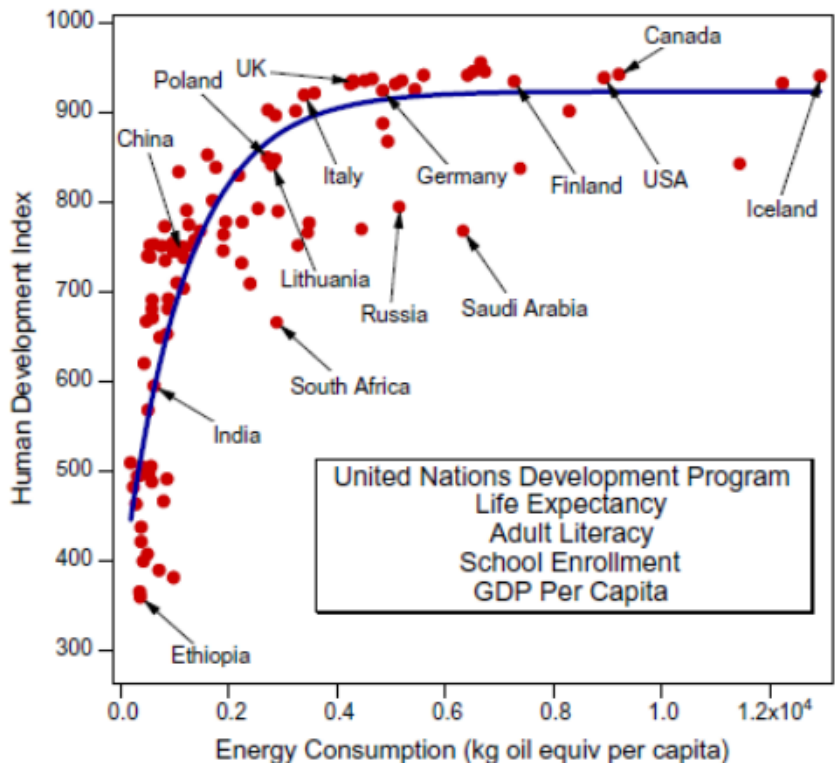

Figure 3 - The United Nations links Energy per capita to Quality of Life.

REME will reduce environmental degradation compared to our current use of mining to provide our construction materials. It will also increase equity because of the availability of sun, water and air, inputs to the REME economy, in the global south. REME can be done at any scale and is standalone and thus can be implemented regionally without the need for expensive infrastructure that characterizes our current fossil fuel economy. There is also a positive equity impact of the capital savings for future new long-distance transportation commercial infrastructure (e.g., pipelines and ships and ports) if that capital instead is used to build schools, hospitals and housing as well as local infrastructure, especially in the global south.

Finally, while REME is based upon $CO_2$ from the air, there needs to be a transition period to reduced use of fossil fuels. In this analysis, $CO_2$ can come from capturing $CO_2$ from flue gas to reduce emissions or from carbon dioxide in the air. For climate change protection and for a sustainable economy in the future the source of carbon needs to come from air like it does in natural systems[4].

## Energy Use

The most challenging of the technologies needed for REME to sequester carbon in materials is the conversion of $CO_2$ into carbon fibers, nanotubes, or graphene as well as synthetic aggregate. The conversion to carbon fiber has until recently been dismissed because of the high-energy use needed to split the $CO_2$ bond. Or as many experts assert, it makes no sense to reverse the combustion process but that is of course, exactly what nature does. It will be shown below that carbon-based construction materials can be produced using less of a fraction of our total energy than the 11% for building materials used in 2019[5].

Favoring the reduced need for energy for materials of construction is the enhanced properties of carbon fiber per weight, in particular about a factor of 15 greater strength-to-weight ratios than steel and a factor of about 4-5 for aluminum [6][7] The strength-to-weight ratio of carbon fiber is over 150 greater than concrete.

In general, compared with conventional concrete, steel, and aluminum, the production of carbon fiber is much costlier than mining the inputs to conventional materials of construction. But the costs of transforming their ores into usable materials for construction is more energy-intensive per structure. The approach used for comparison is embodied energy per kg, which is about 30 MJ/kilogram for steel, 200 MJ/tonne for aluminum, 1.5 MJ/tonne for concrete[8], and for the less developed carbon fiber at scale with heat management, 100 MJ/tonne or lower[9] . Thus, using the ratios of strength to weight one has for an equivalent structure 450MJ/tonne for steel, 800 MJ/tonne for aluminum and 225 for concrete. This makes using carbon fiber less energy-intensive for making our structures than the materials currently used. Carbon fiber reinforced concrete can offer advantages for specific uses.

Another major sequestration opportunity is synthetic aggregate because 440 kg of $CO_2$ is sequestered for each tonne of synthetic concrete. Synthetic aggregate can be produced with less energy use and reduced cost to make concrete than the processes used today[10].

 Note that the energy to capture CO2 is 5 MJ/tonne for direct air capture and less for emission reduction. Most processes use low-temperature heat that is currently wasted in the production of electricity and high temperature industrial processes. It can be obtained via cogeneration with renewable energy sources or carbon fiber production, which emits a lot of heat. This increases overall energy efficiency and reduces the cost and need for primary energy for making synthetic materials. Specifically, heat integration with electrical energy production via renewable energy sources like nuclear or concentrated solar thermal or fusion in the future offers opportunities to reduce the amount of additional primary energy production needed.[11] The above analysis shows that making synthetic materials will take less energy in a REME economy compared to the 11% used in the current extractive materials economy.

## Cost

Carbon fiber has experienced a cost reduction of 3 between 2012 and 2020 as its use grows, reaching under $10 dollars/ Kg or 10,000 per tonne[12]. This is a 15% learning rate. Solar panels experienced a 400-fold reduction in cost over 50 years as their use grew since their first use in space. This is a 13% learning rate. As the rate for carbon fiber production continues to increase, possibly at an enhanced rate in time because of the enhanced economic growth rate, one can expect costs of $1000 per tonne or less. The cost for steel, for example, varies greatly, but for structural uses, about $1000 per tonne, with stainless steel being much higher and other products lower. For aluminum, the cost is about $2000 per tonne. Those industries will make incremental cost reductions in processing their ores but will likely be offset by the increased ore costs as supplies become limited. For example, iron ore known reserves can make 85 billion tonnes of steel, and the world use currently is about 2 billion tonnes per year [13] [14]. The same can be said about aluminum and even aggregate, which requires extra transportation as sources near cities are depleted[15].

Current uses in the aerospace industry of carbon fiber are at much higher costs per tonne, as high as $200,000[16]. Thus, carbon fiber has applications that can support high costs while it reduces its cost as it scales as will carbon capture and other REME technologies. It will be economically feasible at the large scale needed. A sustainable synthetic materials economy (REME) is needed even if there is no climate threat because of the limited supply, energy intensity, and environmental damage of extracting natural resources. So, cost is not a barrier to sequestering carbon in synthetic materials. It will be profitable at scale while building the new sustainable infrastructure needed for increased equity and climate change protection.

# Paris Targets

The remaining and very critical question is whether there will be enough demand for materials to sequester the amount of carbon needed for climate protection. There are other ways to sequester carbon, like biochar and our chemicals, and even desalinated water treatment. There are natural land use changes and afforestation that can sequester carbon. Those should be pursued largely for environmental reasons since they cannot scale to the level needed for prevent a climate disaster so they will ignore them for this zeroth order assessment. The amount of carbon we can store in structural materials, CCUS in gigatonnes/yr, is determined by the amount of materials of construction we use each year, MP in gigatonnes/yr, times their carbon intensity, CI, where CI is the fraction of the weight that is carbon.

$$(1) \quad CCUS = MP \times CI$$

The world produced over 35 gigatonnes of construction materials in 2009 (see graph below). If our future buildings are composed of carbon, one needs to adjust for the reduced weight of a carbon fiber composite compared to steel, aluminum, and concrete. If synthetic aggregate is used its ability to sequester carbon its carbon intensity is about 17%. For this back-of-the-envelope calculation, one can consider a conservative factor of about 3.5 for the average weight reduction to meet our needs given the large role of synthetic aggregate. So, the equivalent amount of carbon would be about 10 gigatonnes in 2009.

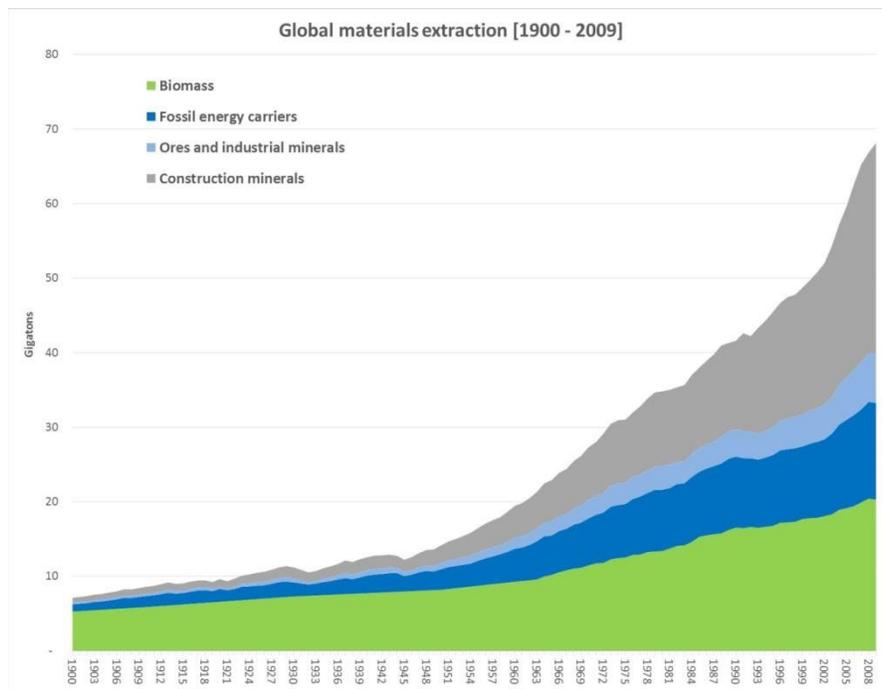

Fig. 2. Global Material Extraction (1900-2009)[17]

Recent use of extractive materials has grown to over 100 gigatonnes, with structural materials accounting for most of the increase[18]. Here, we will use 50 gigatonnes for extractive materials. We will use a factor of 10 growth in materials as a conservative estimate from a factor of 20 in Gross Global Domestic Product in 50 years resulting from a 6% growth rate[19].

Using 500 gigatonnes and if it was all carbon the weight of the materials used would be 150 gigatonnes. So, a carbon intensity CI of only 7% is needed for ten gigatonnes or 36 gigatonnes of annual $CO_2$ removal in fifty years. Equation 1 makes clear that as the use of materials increases the carbon intensity needed is reduced for the CCUS needed for climate change. This is a simple example of the positive feedback in REME: the more materials you use, the more economic growth, greater equity, and enhanced climate protection.

There is today, over 200,000 tonnes per year of carbon fiber produced[20]. Using the doubling capacity every two years it would take about 32 years to develop 10 gigatonne capacity. However, synthetic aggregate will reduce the amount of carbon fiber needed. The 16 doublings would, with a 15% learning rate, reduce the cost by about 15 for carbon fiber, resulting in a synthetic materials economy being both less energy intensive and less costly to the global economy.

## Conclusion

It is noteworthy that in REME, the more energy we use, the more synthetic materials we use, the better it is for economic growth, energy security, equity, and climate protection[21]. Because of its economic viability and social benefits, REME is not a cost to the global economy but enables and contributes to global prosperity. Realizing the value of carbon-based materials for construction will be helped by innovations that reduce costs and improve performance. This in turn, will increase the carbon intensity of our construction materials. It makes it possible to sequester the carbon needed to meet the Paris targets. REME both increases economic growth and addresses a global threat that warrants a global mobilization even greater than the US experienced in World War II where the growth rate was 17% per year. The use of energy and materials based upon a 6% growth rate has to be viewed as conservative when the world mobilizes to address the threat of climate change. It certainly makes extrapolations of global growth of 3% a disaster for equity and the climate threats we face. The positive feedback should enable a global accord on an equitable climate effort because it changes IPCC discussions from who pays for the effort to who benefits for addressing our challenges. We have the capability to address the challenges we face. All we have to do is decide to mobilize to use it!